\documentclass{ws-ijmpa}

\begin{document}

\markboth{VIP Collaboration}
{VIP: a search for a violation of Pauli Exclusion Principle}

%%%%%%%%%%%%%%%%%%%%% Publisher's Area please ignore %%%%%%%%%%%%%%%
%
\catchline{}{}{}{}{}
%
%%%%%%%%%%%%%%%%%%%%%%%%%%%%%%%%%%%%%%%%%%%%%%%%%%%%%%%%%%%%%%%%%%%%

\title{VIP: AN EXPERIMENT TO SEARCH FOR A VIOLATION OF THE PAULI EXCLUSION PRINCIPLE.}

\author{M. CARGNELLI, T. ISHIWATARI, J. MARTON, E. WIDMANN, J. ZMESKAL }

\address{``Stefan Meyer'' Institute for Subatomic Physics, Boltzmanngasse 3, A-1090 Vienna, Austria}

\author{S. BARTALUCCI, S. BERTOLUCCI, M. CATITTI, C. CURCEANU (PETRASCU), \\S. DI MATTEO, 
C.GUARALDO, M. ILIESCU, D. PIETREANU, D. L. SIRGHI, F. SIRGHI, \\L. SPERANDIO}

\address{Laboratori Nazionali di Frascati dell'INFN, CP 13,
 Via E. Fermi 40, I-00044, Frascati, Italy}

\author{M. LAUBENSTEIN}

\address{Laboratori Nazionali del Gran Sasso, S.S. 17/bis, I-67010 Assergi (AQ), Italy}

\author{E. MILOTTI \footnote{corresponding author, e-mail: milotti@ts.infn.it}}

\address{Dipartimento di Fisica, Universit\`{a} di Trieste and INFN-- Sezione di
Trieste, Via Valerio, 2, I-34127 Trieste, Italy}

\author{M. BRAGADIREANU, T. PONTA}

\address{``Horia Hulubei'' National Institute of Physics and
 Nuclear Engineering, \\ Str. Atomistilor no. 407, P.O. Box MG-6, 
Bucharest - Magurele, Romania}

\author{J.-P. EGGER}

\address{Institut de Physique de l'Universit\'e, CH--2000 Neuch\^atel,
 Switzerland}

\maketitle

\begin{history}
\received{Day Month Year}
\revised{Day Month Year}
\end{history}

\begin{abstract}
The Pauli Exclusion Principle is a basic principle of Quantum Mechanics, and its validity has never been seriously challenged. However, given its fundamental standing, it is very important to check it as thoroughly as possible. Here we describe the VIP (VIolation of the Pauli exclusion principle) experiment, an improved version of the Ramberg and Snow experiment (E. Ramberg and G. Snow, {\it Phys. Lett. B} {\bf 238}, 438 (1990)); VIP has just completed the installation at the Gran Sasso underground laboratory, and aims to test the Pauli Exclusion Principle for electrons with unprecedented accuracy, down to $\beta^2/2 \approx 10^{-30} - 10^{-31}$. We report preliminary experimental results and briefly discuss some of the implications of a possible violation.

\end{abstract}

\ccode{PACS numbers: 11.30.-j; 03.65.-w; 29.30.Kv; 32.30.Rj}

%\pagebreak

\section{Introduction}
\label{intro}

The Pauli Exclusion Principle plays a fundamental role in our understanding of many physical
phenomena, and although it has always been spectacularly confirmed by the number and accuracy of its predictions, its foundation lies deep in the structure of quantum field theory and has defied all attempts to produce a simple proof. 
The Exclusion Principle for electrons was tested to a high level of accuracy by Ramberg and Snow (RS)\cite{RS} in 1989, and this measurement has stood unrivaled for several years. The RS experiment used gas X-ray detectors, however far better X-ray detectors are available  now. Here we describe the VIP experiment, which uses CCD's as X-ray detectors and operates in a very clean environment, and whose aim is at least a 1000-fold reduction of the RS bound on the validity of Pauli Exclusion Principle.

The basic idea of the RS experiment was first suggested by George Snow and was put forward in a paper of Greenberg and Mohapatra\cite{GB1}: if there are electron states with wavefunctions with mixed symmetry, and if some electrons sit in the ground states of some atoms while others are free, then the free electrons may be captured in the ground atomic states, even though the ground (1S) states may already be filled with electron pairs. Such a radiative capture process is accompanied by the emission of X-rays, and because of the anomalous atomic electron configuration these X-rays differ from the characteristic X-rays emitted by the atom. The detection of these anomalous X-rays would indicate that non-Paulian atoms exist and would demonstrate a violation of the Pauli Exclusion Principle.
One might argue that if such mixed symmetry states exist, any anomalous pair in a block of material should already fill the ground states of the atoms in the block and therefore no detection should be expected: George Snow suggested that one could reestablish a non-equilibrium situation simply by injecting fresh electrons with an external source, such as a current source. The current source substitutes the radioactive $\beta$-source in the experiment of Goldhaber and Scharff-Goldhaber\cite{GSG} (as reinterpreted by Reines and Sobel\cite{FRS}), which was actually the first search for anomalous X-rays. 

\begin{figure}[t]
\begin{center}
\centering
\includegraphics[bb= 30 132 449 731, angle=-90, width=4in]{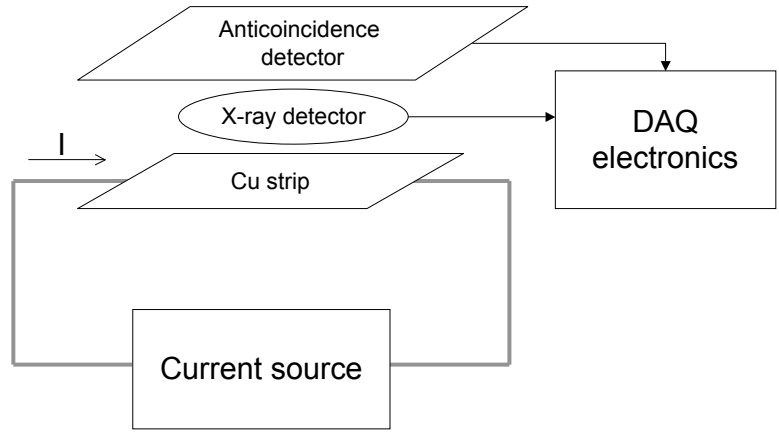}
\vspace*{8pt}
\caption{Sketch of the Ramberg and Snow experimental setup.}
\label{fig1}
\end{center}
\end{figure}

Figure \ref{fig1} shows a simple sketch of the RS experimental layout: a current source injects a current in a copper strip. The copper strip is monitored by an X-ray detector, and an anticoincidence detector rejects spurious events due to cosmic rays. Data acquisition alternates periods with current on and current off, and afterwards the X-ray spectrum obtained with current on is compared with the spectrum with current off. RS found no significant difference between these spectra, and were thus able to set a stringent limit on the probability that electrons actually violate the Pauli Exclusion Principle (they found a Pauli violating probability $\beta^2/2 < 1.7 \cdot 10^{-26}$).

\section{The VIP setup}
\label{setup}

RS used a gas X-ray detector in a noisy environment (Fermilab). Much better soft X-ray detectors are available now, and it is possible to replicate the original RS experiment in low-radiation areas, like the existing underground laboratories. And indeed the idea of the VIP experiment was originated by the 
availability of the excellent DEAR (DAFNE Exotic Atom Research) detector,  just after it had successfully completed its program at the DAFNE collider at LNF-INFN\cite{DEAR}. 
DEAR  used Charge-Coupled Detectors (CCD) to measure exotic atoms (kaonic nitrogen and 
kaonic hydrogen) X-rays transitions. CCDs are almost ideal detectors for X-rays measurement, thanks to their excellent background rejection capability, which is based on the selection of the topology of the events\cite{CCD}, and to their good energy resolution (320 eV FWHM at 8 keV in the present measurement).

\begin{figure}[t]
\begin{center}
\centering
\includegraphics[bb= 30 132 520 731, angle=-90, width=4in]{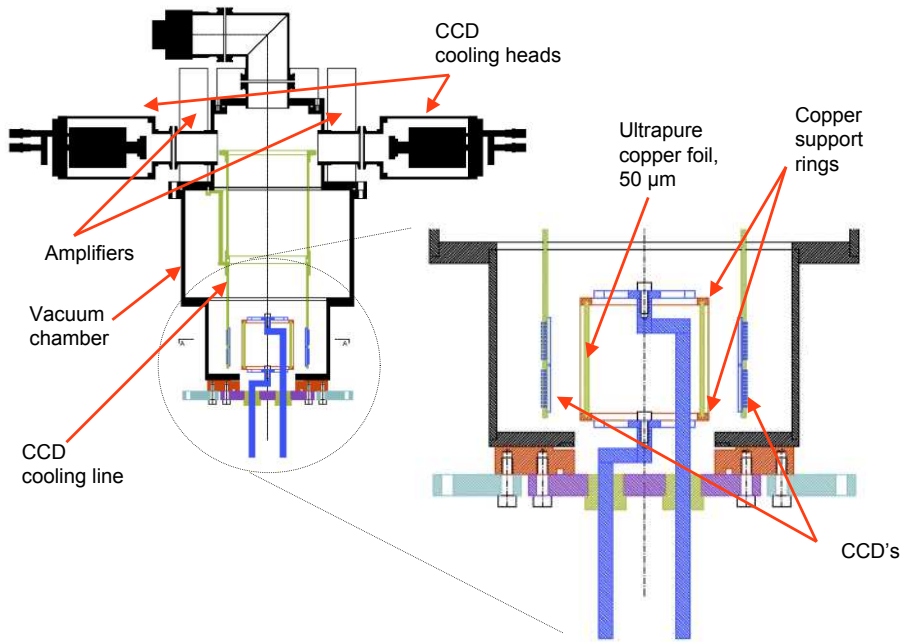}
\vspace*{8pt}
\caption{Layout of the VIP vacuum chamber: the inset shows an enlargement of the copper cylinder and CCD layout.}
\label{fig2}
\end{center}
\end{figure}

The VIP setup  consists of a copper cylinder, which corresponds to the copper strip in the RS experiment, with 45 mm radius, 50 $\mu$m thickness, 88 mm height, surrounded by 
16 equally spaced CCDs of type 55 made by EEV\cite{EEV}. The CCDs are at a distance of 23 mm from the copper cylinder, grouped in units of two chips, one above the other.
The setup is enclosed in a vacuum chamber, and the CCDs are cooled to about 168 K by the use of a cryogenic system, as shown in figure \ref{fig2}. The current flows in the thin cylinder made of ultrapure copper foil at the bottom of the vacuum chamber. The CCD's surround the cylinder and are supported by cooling fingers that project from the cooling heads in the upper part of the chamber. The CCD readout electronics is just behind the cooling fingers; the signals are sent to amplifiers on the top of the chamber. The amplified signals are read out by ADC boards in the data acquisition computer. 

More details on the CCD-55 performance, as well on the analysis method used to reject background events, can be found in reference \refcite{Ishi}.

\section{First results obtained at LNF}
\label{results}

\begin{figure}[t]
\begin{center}
\centering
\includegraphics[angle=-90, width=4in]{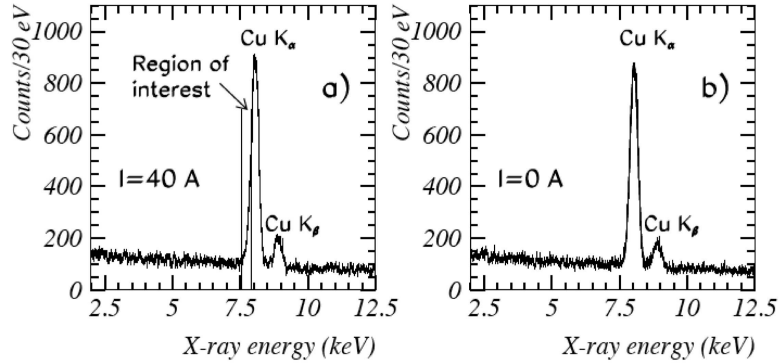}
\vspace*{8pt}
\caption{Energy spectra in the Frascati measurement: a) energy spectrum for the measurement with 
current (I=40 A); b) energy spectrum for the measurement without current.}
\label{fig3}
\end{center}
\end{figure}

The VIP setup is presently taking data in the low-background Gran Sasso underground laboratory of the Italian Institute for Nuclear Physics (I.N.F.N.), however it was first prepared and tested in the Frascati I.N.F.N. laboratories, and measurements were taken there in the period 21 November - 13 December 2005. 
Two types of measurements were performed:
\begin{itemize}
\item{14510 minutes (about 10 days)
of measurements with a 40 A current circulating
in the copper target;} 
\item{14510 minutes of measurements without circulating
current,} 
\end{itemize}
\noindent
where CCD's were read-out every 10 minutes.
The resulting calibrated in energy X-ray spectra are shown in figure \ref{fig3}. These spectra include data from 14 CCD's out of 16, because of noise problems in the remaining 2.
Calibration runs with an iron source and detector checks were done on a daily basis; we estimate that the systematic error in the energy scale at 6 keV is kept below 2 eV.  Both spectra, apart of the continuous background component generated by the cosmic ray background and natural radioactivity, display clear Cu $K_{\alpha}$ and $K_{\beta}$ lines due to X-ray fluorescence also caused by the cosmic ray background and natural radioactivity. No other lines are present and this reflects the careful  choice of the materials used in the setup, as for example the high purity copper and high purity aluminium.
We remark that the copper $K$ lines were not as evident in the original RS setup because of the anticoincidence detector, while CCD's are integrating, non-triggerable detectors.
The region of interest marked in the spectra is where we expect the anomalous X-rays, and it has been calculated with a multiconfiguration Dirac-Fock method with an error estimated as better than 10 eV (see reference  \refcite{DMS} for computational details); the position of the anomalous X-rays should differ from the normal $K_{\alpha}$ transition by about 300 eV (7.729 keV instead of 8.040 keV).

\begin{figure}[t]
\begin{center}
\centering
\includegraphics[angle=-90, width=4in]{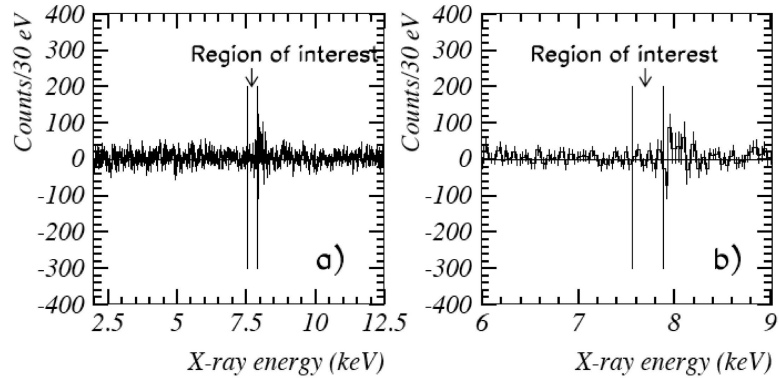}
\vspace*{8pt}
\caption{Subtracted energy spectra in the Frascati measurement, current minus no-current, giving the
limit on PEP violation for electrons: a) whole energy range; b) expanded view 
in the region of interest (7.564 - 7.894 keV).  
No evidence for a peak in the  region of interest is found. }
\label{fig4}
\end{center}
\end{figure}

The subtracted spectrum is shown in Figure 3 a) (whole energy scale) and
b) (a zoom on the region of interest).
Notice that the subtracted spectrum is normalized to zero
within statistical error, and is structureless. This not only yields an upper bound for
a violation of the Pauli Exclusion Principle for electrons, but also confirms the correctness of
the energy calibration procedure.

The Pauli Exclusion Principle is deeply embedded in quantum theory and it is very difficult to put a violation in a proper theoretical frame. Here we use the notation of  Ignatiev and Kuzmin \cite{IK}, which has been incorporated in the paper of Greenberg and Mohapatra\cite{GB1}: even though the model of Ignatiev and Kuzmin has been later shown to be incompatible with quantum field theory\cite{Govorkov}, the parameter $\beta$  that measures the degree of PEP violation in the Ignatiev and Kuzmin model has stuck and is still found in the literature, also because it is easy to show that it is related to the parameter $q$ of quon theory, by the relation\cite{green}: $(1+q)/2 = \beta ^2/2$ (in quon theory, $-1 \le q \le 1$, 
where $q=-1$  corresponds to fermions and $q=1$ corresponds to bosons, so that here  $q$ must be close to -1 and $(1+q)/2$ must be very small, because we are dealing with electrons).
Moreover, we used this parametrization for an easy comparison of our results with the RS experiment, which utilized the same formalism.
To determine the experimental limit on $\beta^{2}/2$ from our data, we used the same arguments of Ramberg and Snow: see references \refcite{RS} and \refcite{PLA} for details of the analysis. Finally we find\cite{PLA}:
\[
\frac{\beta^{2}}{2} \leq 4.5 \times 10^{-28} \mathrm{ ~at \; 99.7 \% \; CL.}
\]
Thus in this first run in the Frascati laboratory we have already improved the limit obtained by Ramberg and Snow by a factor about 40.

\section{Discussion}
\label{discussion}

The Pauli Exclusion Principle is an almost direct consequence of the spin-statistics connection which was first proved by Pauli in 1940\cite{Pauli}. Pauli's proof is not easy to follow, and it was soon followed by other arguments that tried to give the theorem a more obvious basis. In particular the paper by  
L\"uders and Zumino\cite{LZ} states very clearly the postulates that must be fulfilled by a field theory for the spin-statistics connection to hold:
\begin{itemize}
\item the theory is invariant with respect to the proper inhomogeneous Lorentz group (this includes translations, but does not include reflections);
\item two operators of the same field at points separated by a spacelike interval either commute or anticommute;
\item the vacuum is the state of lowest energy;
\item the metric of the Hilbert space is positive definite;
\item the vacuum is not identically annihilated by a field.
\end{itemize}
From these postulates it follows that (pseudo)scalar fields commute and spinor fields anticommute, and if we hypothesize a violation of the Pauli Exclusion Principle and assume the validity of the field theory description of the subnuclear world, this implies that some of these postulates are necessarily violated.
In particular there could be a violation of locality, and if this is the case it is possible to use Medvedev's stochastic intepretation of ambiguous statistics\cite{med} (in which electrons are normally fermions, but behave as bosons during a small -- and random -- fraction of time) and the VIP result reported above to find an upper bound for the length scale of a possible locality violation: $\ell < 1.35\cdot 10^{-19}$ m. 

The ongoing measurement in the Gran Sasso--I.N.F.N. underground laboratory (LNGS), uses higher integrated currents, and from preliminary tests, it appears that the X-ray background in the LNGS environment is a factor 10-100 lower than in the Frascati Laboratories. The experiment runs unguarded and is monitored remotely from the Frascati Laboratories: in the long run it needs only a minimal maintenance. 

A VIP measurement of two years (one with current, one without current) at LNGS, has started taking data in spring 2006, and we estimate that this will bring the limit on a violation of the Pauli Exclusion Principle for electrons into  the 10$^{-30}$ -- 10$^{-31}$ range, and the upper bound for the length scale of a possible locality violation into the range 10$^{-21}$ m -- 10$^{-22}$ m.

%\begin{thebibliography}{000} %for 3 digits

\end{document}